\documentclass[12pt]{article}
\usepackage{graphicx,indent,comment,cite}

\makeatletter
\long\def\@makefntext#1{\parindent 1em\noindent
\hbox to 2em{\hss$^{\@thefnmark}$~}%
\@tempdima\columnwidth\advance\@tempdima-2em
\parbox[t]{\@tempdima}{#1}}
\makeatother

\begin{document}

\begin{center}
\textbf{\large Melting of 2D Coulomb clusters in dusty plasmas\\}
\vspace{10pt}
R. Ichiki$^a$\footnote{Present address:
Department of Electronic Engineering, Tohoku University,
Sendai 980-8579, Japan
Electronic mail: ryu@ecei.tohoku.ac.jp},
Y. Ivanov$^b$, M. Wolter$^b$, Y. Kawai$^a$,
and A. Melzer$^b$\\
\vspace{10pt}
\textit{
$^a$Interdisciplinary Graduate School of Engineering Sciences,
Kyushu University\\
Kasuga, Fukuoka 816-8580, Japan\\
$^b$Institut f\"ur Physik,
Ernst-Moritz-Arndt-Universit\"at Greifswald\\
17489 Greifswald, Germany\\}
\end{center}

\begin{indentation}{10mm}{10mm}
\noindent \textbf{Abstract}

The melting of 2D dust clusters caused by one additional particle in
the lower layer has experimentally been observed
to undergo a two-step transition,
which divides the phase of the cluster into three stages.
The first transition is a jump of the dust
kinetic energy due to the onset of an instability
of the lower-layer particle,
shifting the cluster from an ordinary to a hot crystalline state.
The second transition is
the actual phase transition into a liquid state,
which occurs at a decisively lower gas pressure.
The detailed dynamical properties of the system during the transition
were determined in terms of the normal mode analysis.
\end{indentation}
\vspace{10mm}

Dusty plasmas are ideal systems to study the dynamics of crystalline,
fluid and gas-like charged particle systems
since the spatial and temporal scales are perfectly suited for
direct observation by video cameras.
The particles are highly negatively charged due to the
continuous inflow of plasma electrons and ions.
Micrometer-sized particles usually attain
charge numbers $Z$ of the order of 10 000.
Due to these high charges the dust particles arrange
in ordered Coulomb lattices.
Finite dust clusters consist of a small number of dust particles
$N$ immersed in a gaseous plasma environment.
Dust clusters in two dimensions (2D) are formed by trapping
the dust particles in the sheath above a bowl-shaped electrode.
Vertically the particles are strongly confined due to the balance
of electric field force and gravity.
A much weaker horizontal confinement is provided by the distorted
equipotential lines of the curved electrode.
2D clusters arrange in concentric rings (see e.g.\ Refs.~\citen{Juan98}
and \citen{Melzer03}).

In this paper,
we will focus on the dynamics of 2D clusters under a phase transition from
the solid to the liquid state.
Phase transitions can be driven by reducing the gas pressure in
multi-layer dust systems~\cite{Melzer96,Thomas96}.
This is due to an oscillatory instability of the lower layer
particles which is excited by the ion streaming motion in the
sheath~\cite{Schweigert98}.
The oscillations heat the dust particles which leads to
the melting of the ordered dust system.
Here, the full dynamical properties of a dust cluster
with $N = 42$ particles is determined during the phase transition.
The dynamics are derived in form of the mode spectrum of
the $2N$ cluster modes.
\begin{figure}
\begin{minipage}{0.5\linewidth}
\includegraphics[width=80mm]{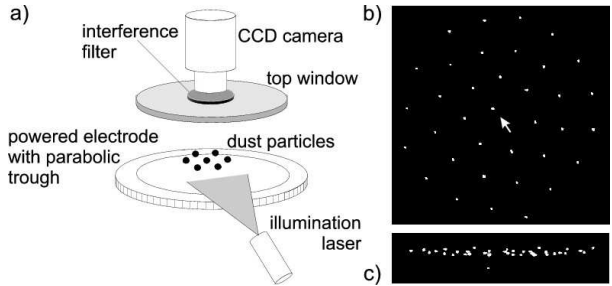}
\end{minipage}
\hspace{0.05\linewidth}
\begin{minipage}{0.45\linewidth}
\caption{\label{fig:schematic}
\small
(a) Scheme of the experimental setup.
(b,c) Snapshots of a cluster with $N = 36$ upper particles
and a single lower particle.
(b) Top view. The arbrow indicates the particle under which
the lower particle is located.
(c) Side view. The side view camera is slightly tilted with respect to the
cluster plane, so the upper layer appears as an elliptical disc.}
\end{minipage}
\end{figure}
To pinpoint the melting transition to the instability of
the lower layer particles a dust cluster with only
a single lower layer particle was prepared (see Fig.~\ref{fig:schematic}).
The dynamic properties of the cluster were directly obtained
from the thermal motion of the upper layer particles during
the phase transition.
This technique is described in detail in Ref.~\citen{Melzer03}.
The analyzed video sequences cover 41 s at a frame rate of
50 frames per second.

\begin{figure}
\begin{minipage}{0.6\linewidth}
\includegraphics[width=\linewidth]{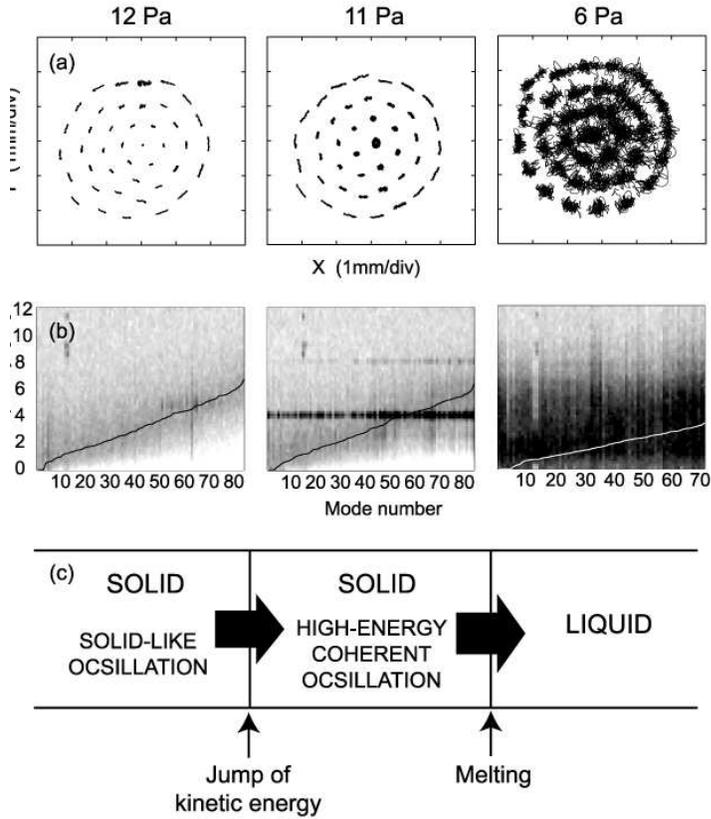}
\end{minipage}
\hspace{0.05\linewidth}
\begin{minipage}{0.35\linewidth}
\caption{\label{fig:data}
\small
Melting transition of the dust cluster with
decreasing gas pressure.
(a) Trajectories of the cluster particles.
(b) Gray-scale power spectra of the normal mode oscillations of the cluster.
Darker colors correspond to higher power density.
The theoretical mode frequencies for a solid state are also indicated
as the black or white solid lines.
(c) Phase diagram including the feature of particle oscillations.}
\end{minipage}
\end{figure}
\begin{figure}
\begin{minipage}{0.4\linewidth}
\includegraphics[width=\linewidth]{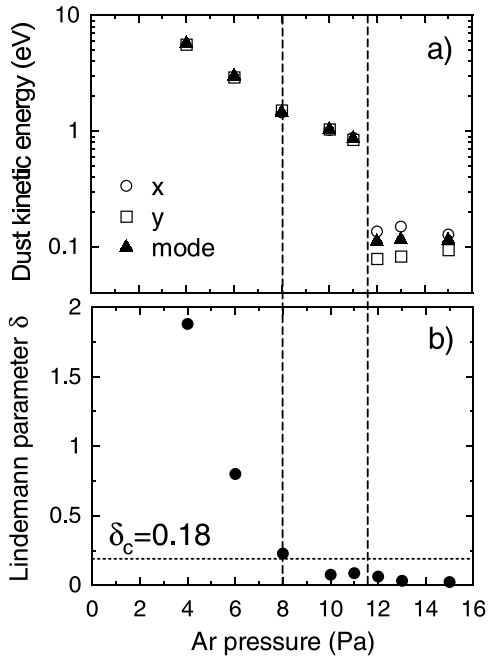}
\end{minipage}
\hspace{0.1\linewidth}
\begin{minipage}{0.5\linewidth}
\caption{\label{fig:energy}
\small
(a) Average kinetic energy of the dust particles
in $x$ and $y$ directions and that obtained from the mode spectra.
(b) Lindemann parameter $\delta$ as a function of Ar pressure.}
\end{minipage}
\end{figure}
Figure \ref{fig:data} shows the dynamical properties of the cluster
during the phase transition for 3 representative gas pressures
of 12 Pa, 11 Pa and 6 Pa.
From the particle trajectories [Fig.~\ref{fig:data}(a)] at 12 and 11 Pa
the cluster is seen to be well ordered whereas at 6 Pa
many changes in equilibrium positions can be identified.
Moreover,
at 11 Pa the particles are found to exhibit
an oscillatory motion which are notable from the circular particle
trajectories,
especially for that central particle under which the
lower particle is located
[see the arrow in Fig.~\ref{fig:schematic}(b)].

In fact, the cluster changes from the solid state (at 12 Pa) to
an intermediate state (at 11 Pa) and, finally,
to the liquid state (at 6 Pa).
As shown in Fig.~\ref{fig:energy},
the change of the cluster dynamics is accompanied by a change
of the global parameters that describe the dust system,
namely the dust kinetic energy and the Lindemann order parameter $\delta$.
The kinetic energy suddenly increases from about 0.1 eV
at 12 Pa to 1 eV at 11 Pa and even further to 5 eV at 6 Pa.
This energy jump corresponds to
the onset of the instability which effectively heats the dust system.
The Lindemann parameter $\delta$ that describes the root-mean-square
excursions of the particle from the equilibrium positions stays
at a very low level down to gas pressures of 8 Pa.
This indicates strong order. Below 8 Pa, $\delta$ suddenly jumps
to large values which is a clear sign of melting.
That is to say,
the melting involves two-stages of transitions:
the jump of the dust energy and the actual melting,
which divide the phase of the cluster into three states:
the solid, intermediate, and liquid states.

The analysis of the normal mode spectra provides
detailed characteristics of these three states,
especially of the intermediate state.
Figure \ref{fig:data}(b) shows the mode spectra obtained
from the thermal motion of the particles together
with the theoretical mode frequencies.
In the solid state at 12 Pa the mode spectrum closely follows
the expected mode frequencies.
In the intermediate state at 11 Pa the situation is drastically different.
From the intense horizontal band a dominating oscillatory motion at
a frequency of about 4 Hz is observable.
This is exactly the frequency of the unstable oscillation
of the lower particle that sets in at exactly this gas pressure.
It is somewhat surprising that the single lower particle dominates
all modes of the cluster.
The second harmonic of the unstable oscillations at 8 Hz is
also detectable.
A closer inspection of the modes reveals that the expected mode structure
of the solid state is also faintly visible in the spectrum.
In contrast, in the liquid state at 6 Pa the spectrum is broad
for all modes and does not at all
resemble the mode spectrum of the solid state.
Figure \ref{fig:data}(c) illustrates the phase diagram of the cluster.

The dynamic behavior is even more clearly seen in
the mode-integrated spectra
or the power spectral density of the entire system,
but the details will be reported elsewhere.

Concluding, three different phases of the dust cluster have been identified.
At high gas pressures, the cluster is in a solid state with high order,
low dust kinetic energy and a solidlike mode spectrum.
In the intermediate phase, the cluster is dominated
by the unstable oscillations.
The particles are heated, but the system is still in an ordered state.
Thus, this state can be characterized as a hot, oscillating crystal.
At low gas pressures the cluster goes to the liquid phase with even
hotter instability-heated particles, low order and a broad mode spectrum.
The analysis of the cluster dynamics thus demonstrates
that the heating can be definitely attributed to the heating
by the oscillatory instability of the lower layer particle.

\end{document}